\documentstyle[11pt,newpasp,twoside,epsf]{article}
\def\edcomment#1{\iffalse\marginpar{\raggedright\sl#1\/}\else\relax\fi}
\reversemarginpar
\begin{document}
\title{Density profiles and clustering of dark halos and clusters of galaxies}
 \author{Yasushi Suto} 
\affil{Department of Physics, The University of Tokyo, Tokyo 113-0033, Japan.}

\begin{abstract}
Density profiles of cosmological virialized systems, or dark halos, have
recently attracted much attention. I first present a brief historical
review of numerical simulations to quantify the halo density
profiles. Then I describe the latest results on the universal density
profile and their observational confrontation. Finally I discuss a
clustering model of those halos with particular emphasis on the
cosmological light-cone effect.
\end{abstract}

\section{Introduction}

The key assumption underlying the standard picture of structure
formation is that the luminous objects form in a gravitational potential
of dark matter halos.  Therefore, a detailed description of halo density
profiles as well as of their clustering properties is the most basic
step toward constructing the formation and evolution of galaxies and
clusters.

More specifically, the importance of the detailed studies of density
profiles of dark halos is two-fold:

\begin{description}
 \item[(i) theoretical interest;]\hfill\par 

What is the final (quasi-)equilibrium state of cosmological
self-gravitating systems (as long as the energy dissipation is
neglected) ?  One may easily think of two quite distinct, but equally
plausible, possibilities;
\begin{enumerate}
\item[(A)] the systems reach a certain universal distribution
	   which is independent of the cosmological initial condition.
\item[(B)] the systems somehow keep the memory of the cosmological
	   initial condition even at the highly nonlinear regime.
\end{enumerate}
The singular isothermal sphere:
\begin{eqnarray}
 \rho(r) = \frac{\sigma^2}{2\pi G} \frac{1}{r^2}
\end{eqnarray}
may be a reasonable possibility along the line of the idea (A).  As a
matter of fact, quite often I meet people who argue that the idea (B) is
unlikely because of the strong nonlinear nature of the gravitation. In
such an occasion, I present an example of the well-known stable
clustering solution for the nonlinear two-point correlation function
	    (Davis \& Peebles 1977):
\begin{eqnarray}
P_{\rm mass}(k) &\propto& k^n 
\quad \rightarrow\quad 
\xi_{\rm mass}(r) \propto r^{-3(n+3)/(n+5)} .
\end{eqnarray}
This provides a good specific case that the cosmological initial
condition is not erased and imprinted even in the strongly nonlinear
behavior as is well confirmed by later numerical simulations (e.g., Suto
1993; Suginohara et al. 2001).  Of course the answer to the final state
of cosmological self-gravitating systems may not be unique since it
should really depend on the specifics and scales of the systems under
consideration. For instance, it is clearly hopeless to extract any
meaningful cosmological information from the precise data on the orbit
of the earth around the Sun; all the initial memory should have been
lost due to the strongly nonlinear and chaotic nature of the
gravitation.  Nevertheless the final state of dark matter halos
corresponding to galaxy- and cluster-scales is a well-defined problem
which may be reliably answered with the current high-resolution
numerical simulations independently of the physical intuition.

\item[(ii) practical application;]\hfill\par 

Whether or not the density profiles of dark halos keep the cosmological
initial memory, the quantitative (empirical) prediction of the profile
for a given set of cosmological parameters has several profound
astrophysical implications including the rotation curve of spiral
galaxies, reconstruction of the mass distribution from the weak-lensing,
and X-ray and SZ observations of galaxy clusters. In particular, the
confrontation of those testable predictions against the accurate
observational data may even challenge the cold dark matter paradigm
itself.
\end{description}

In this article, first I will review the summary of the past studies of
the density profiles of dark matter halos, and then present some
applications of those results.

\section{Previous work on the density profiles of dark matter halos}

The study of the density profiles of cosmological self-gravitating
systems or dark halos has a long history. Table 2 is my personal (and
thus incomplete) summary of some past work that I recognize (or at least
I have really read those papers).

Hoffman \& Shaham (1985) argued, on the basis of the secondary infall
model (Gunn \& Gott 1972), that the density profile around peaks is
given by $\rho \propto r^{-3(n+3)/(n+4)}$ where $n \equiv d \ln P(k)/d
\ln k$ is the spectral index of the density fluctuation field. The
resulting power index $-3(n+3)/(n+4)$ is close to, but slightly differ
from, that for the stable clustering solution for the nonlinear
two-point correlation function $-3(n+3)/(n+5)$. In any case this result
implies that the density profile of dark halos {\it does} keep the
initial memory.

Many subsequent numerical simulations seem to have confirmed the
prediction of Hoffman \& Shaham (1985). While some authors have reported
the existence of a kind of universal profile, they were not able to
specify its functional form. In this respect, the proposal of the
specific universal density profile by Navarro, Frenk \& White (1995,
1996, 1997) was really a breakthrough. They suggested that all simulated
density profiles can be well fitted to the following simple model (now
referred to as the NFW profile):
\begin{equation}
\label{eq:nfw}
\rho(r) \propto {1 \over (r/r_{\rm s})(1+r/r_{\rm s})^2}
\end{equation}
by an appropriate choice of the scaling radius $r_{\rm s}=r_{\rm s}(M)$
as a function of the halo mass $M$.  Considering the previous conclusion
from numerical simulations, their claim was quite surprising and, in
some sense, contradictory. It turned out that most simulations prior to
their discovery did not have sufficient mass-resolution to reliably
represent the deep central gravitational potential, which gives rise to
an artificial central core. This presents a good example that the
quantitative difference leads to the qualitatively different
interpretation of the physical phenomena; now the density profile of
dark halos seems to lose the initial memory, and thus the answer to the
final (quasi-)equilibrium state of cosmological self-gravitating systems
changes even in a qualitative way; from the possibility (A) to (B)
mentioned in the previous section !

\begin{table}[h]
\caption{The stable solution and the density profile around peaks.}
\begin{center}
\begin{tabular}{ccc}
$n$ & $\xi_{\rm stable} \propto r^{-3(n+3)/(n+5)}$  
& $\rho \propto r^{-3(n+3)/(n+4)}$ \\
\hline 
0 & $r^{-1.8}$ & $r^{-2.25}$ \\
-1 & $r^{-1.5}$ & $r^{-2}$ \\
-2 & $r^{-1}$ & $r^{-1.5}$ \\
\end{tabular} 
\end{center}
\end{table}

If this change of the qualitative conclusion (from Hoffman \& Shaham to
NFW) is really ascribed to the mass-resolution of the numerical
simulations, one should naturally wonder whether or not current
numerical simulations are sufficiently reliable for the present
problem. This motivated many people to examine the convergence of the
profile, in particular, its inner power-law index, with significantly
higher--resolution simulations. Fukushige \& Makino (1997) are the first
to claim that the inner slope of density halos is much steeper than the
NFW value. Their claim was confirmed by a series of systematic studies
by Moore et al. (1998, 1999), and the current consensus among most (but
not all) numerical simulators in this field is the {\it converged}
profile is given by
\begin{equation}
\label{eq:universal}
\rho(r) \propto {1 \over (r/r_{\rm s})^{\alpha}(1+r/r_{\rm s})^{3-\alpha}}
\end{equation}
with $\alpha \approx 1.5$ rather than the NFW value, $\alpha=1$,
(e.g., Fukushige \& Makino 2001).  In most cases this difference is fairly
minor and the NFW profile still provides a good empirical approximation
to the dark matter halo (Jing \& Suto 2002).

I just reproduce two figures from my collaborative work (Jing \& Suto
2000) just to illustrate the validity of the universal density profile
(eq.[4]) despite the fact that their projected images look very diverse
rather than universal.  It is hard to imagine the regularity of the
averaged profiles plotted in Figure 2 behind the apparent diversity
quite visible in Figure 1.

\begin{table}[h]
\caption{Past work on the density profiles of dark matter halos.
\label{tab:halopaper}}
\begin{center}
\begin{tabular}{|cl|}
\hline 
1970 & Peebles \\
 & $N=300$ simulation to reproduce the profile of the
Coma cluster\\
1977 & Gunn \\
 & predicted $\rho \propto r^{-9/4}$ on the basis of secondary
 infall model\\
1985 & Hoffman \& Shaham \\
 & density profiles around the peak: $\rho \propto r^{-3(n+3)/(n+4)}$ \\
1986 & Quinn, Salmon \& Zurek \\
 & confirmed the Hoffman \& Shaham prediction 
($N\approx 1000$ simulation)\\
1987 & West, Dekel \& Oemler \\
& suggested of the existence of a kind of universal density profile\\
& with $N\approx 4000$ simulations. \\
1988 & Frenk, White, Davis \& Efstathiou \\
& reproduced the galactic rotation  curve \\
& in the SCDM model with $N=32^3$ simulations \\
1990 & Hernquist \\
& proposed an analytic model for density profiles of elliptical
 galaxies\\
& $\rho (r) = (Ma/2\pi)r^{-1}(r+a)^{-3}$\\
1994 & Crone, Evrard, \& Richstone \\
& suggested a central cusp (not a core) with $N=64^3$ simulations \\
1995 & Navarro, Frenk \& White \\
& discovered a universality in the halo density profile \\
& from SPH simulations in SCDM,  $N\approx 6000$ per halo\\
& $\rho (r) = 7500 \bar\rho(r/0.2r_{200})^{-1}(1+r/0.2r_{200})^{-2}$\\
1996 & Navarro, Frenk \& White \\
& 19 halos with $N(<r_{\rm vir}) = 5000-10000$ in SCDM \\
& suggested a specific form for the universal density profile \\
1997 & Fukushige \& Makino \\
& found a steeper inner profile with $N=786400$ simulations\\
1997 & Navarro, Frenk \& White \\
& universal density profile in a variety of cosmological models\\
1998 & Syer \& White \\
& repeated merger + tidal disruption model $\rightarrow$
$\rho \propto r^{-3(n+3)/(n+5)}$ \\
1998 & Moore et al. \\
& found a steeper inner profile $\rho \propto r^{-1.4}$\\
& consistent with the previous finding of Fukushige \& Makino\\
1999 & Moore et al. \\
& suggested that  $\rho (r) \propto (r/r_s)^{-1.5}[1+(r/r_s)^{1.5}]^{-1}$ \\
2000 & Jing \& Suto \\
& dependence of the inner slope on the halo mass (controvertial) \\
2002 & Jing \& Suto \\
& triaxial modeling of the universal dark matter density profiles \\
\hline 
\end{tabular} 
\end{center}
\end{table}
\clearpage

\begin{figure}[h]
\begin{center}
\leavevmode\epsfxsize=11.0cm \epsfbox{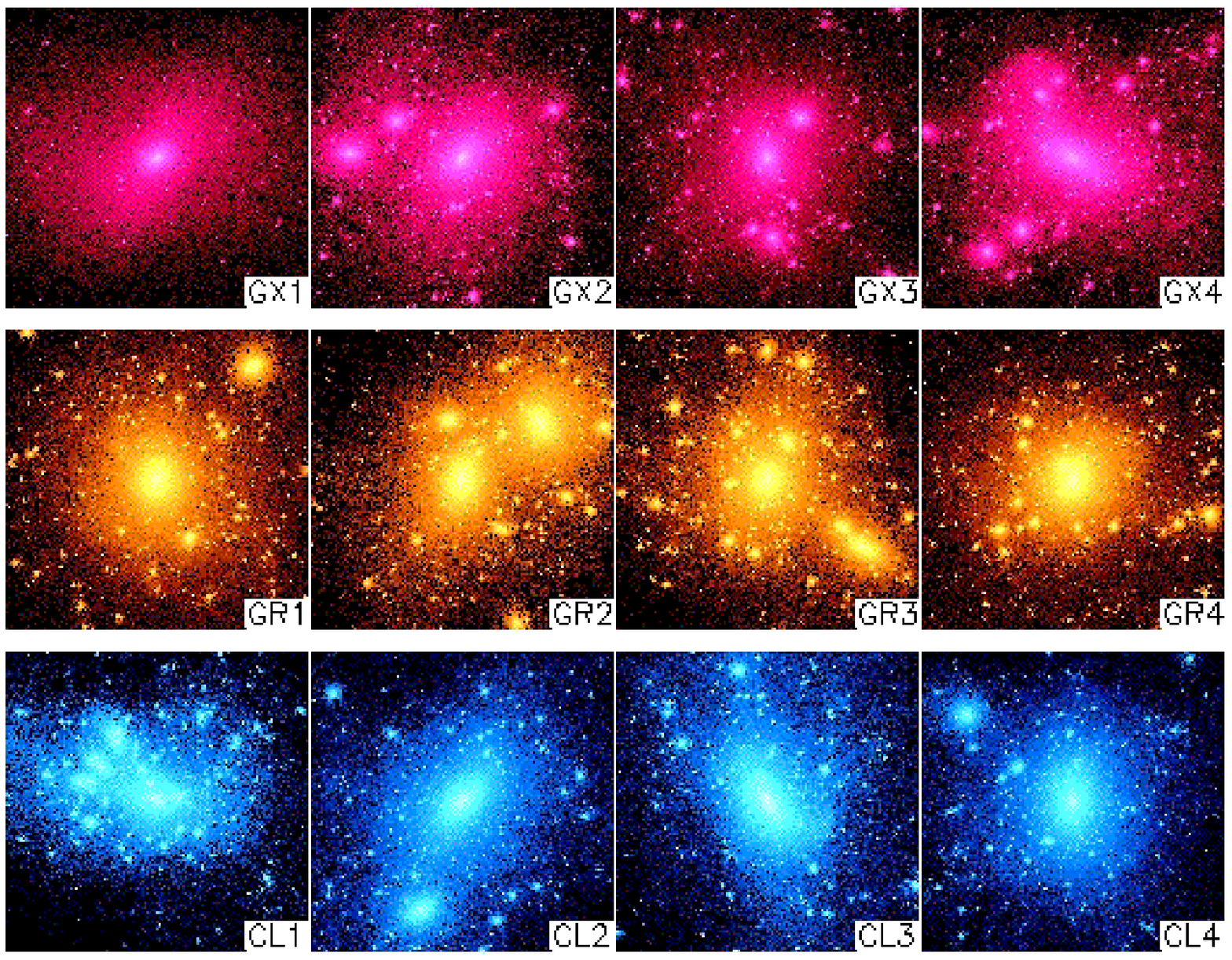} \caption{
Snapshots of the simulated halos at $z=0$.  Top, center and bottom
panels display the halos of galaxy, group and cluster masses (for four
different realizations from left to right), respectively. The size of
each panel corresponds to $2r_{\rm vir}$ of each halo (Jing \& Suto
2000). \label{fig:js00fig1}}
\end{center}
\begin{center}
\leavevmode\epsfxsize=10.0cm \epsfbox{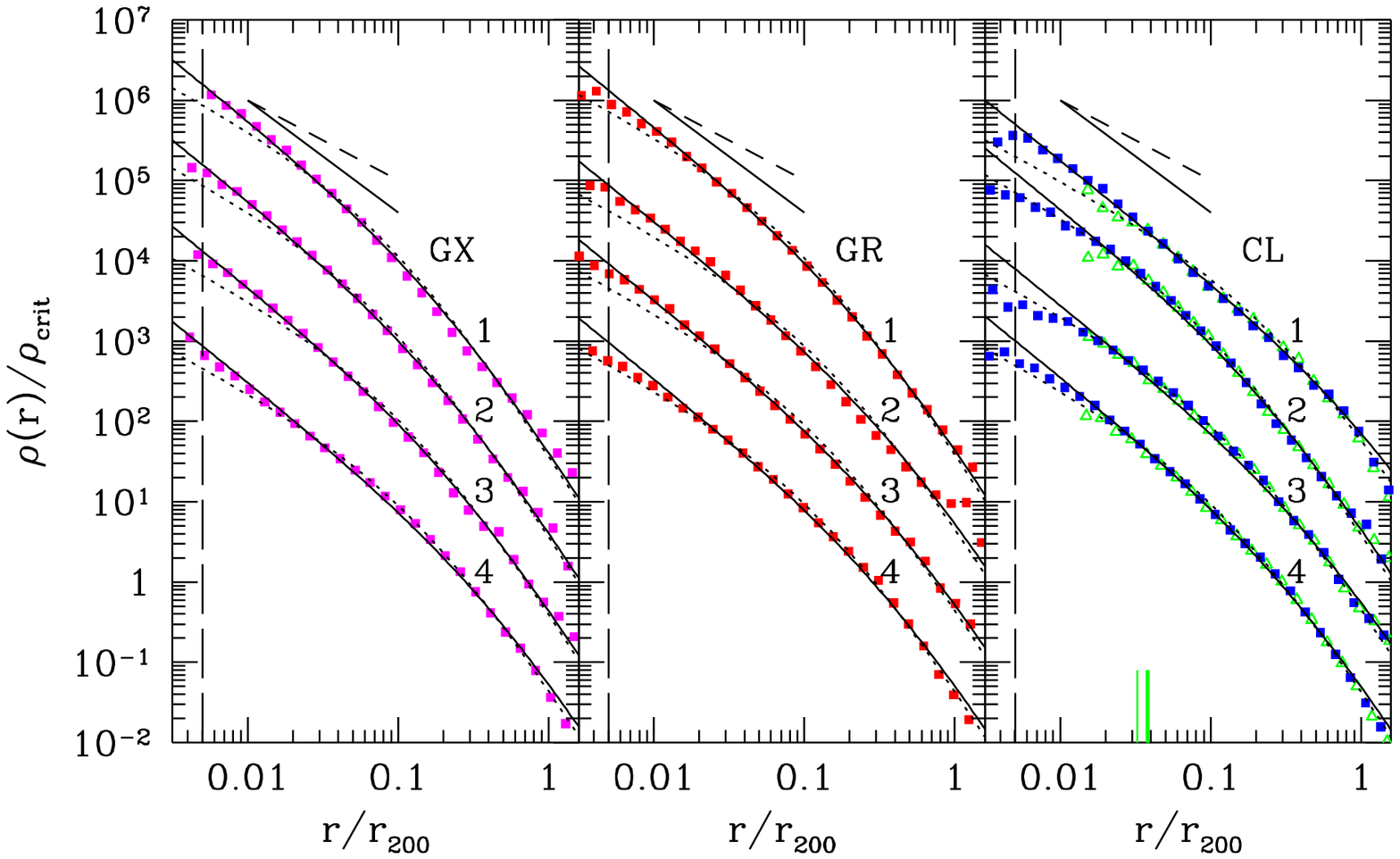} \caption{
Spherically-averaged radial density profiles of the simulated halos of
galaxy ({\it left}), group ({\it middle}), and cluster ({\it right})
masses.  The solid and dotted curves represent fits of $\alpha=1.5$ and
$\alpha=1$ respectively.  For reference, we also show $\rho(r) \propto
r^{-1}$ and $r^{-1.5}$ in dashed and solid lines (Jing \& Suto 2000). 
\label{fig:js00fig2}}
\end{center}
\end{figure}
\clearpage

\section{Observational confrontation of the universal density profile}

\subsection{Mass profile and rotation curve}

The advantage of the NFW profile is that it is completely specified for
a given of cosmological parameters. Therefore this leads to a variety of
testable quantitative predictions. In fact, one can write down all the
necessary formulae in the following concise manner:
\begin{eqnarray}
\rho_{\scriptscriptstyle\rm DM}(r) &=& \frac{\delta_c \rho_{\rm crit}}
 {(r/r_{\rm s})(1+r/r_{\rm s})^2} ,
\end{eqnarray}
where $\rho_{\rm crit}$(z) is the critical density of the universe at $z$
\begin{eqnarray}
\rho_{\rm crit}(z) = \frac{3H_0^2 (1+z)^3}{8\pi G} 
\approx 1.8\times10^{11}(1+z)^3h^2M_\odot/{\rm Mpc}^3 .
\end{eqnarray}
The characteristic halo density contrast $\delta_c$, and the scaling
radius $r_s$ are written in terms of the concentration parameter as
\begin{eqnarray}
\delta_c &=& \frac{\Delta_{\rm vir} \Omega_0}{3}
\frac{c_{\rm vir}^3}{\ln(1+c_{\rm vir})-c_{\rm vir}/(1+c_{\rm vir})}, \\
r_s(M) &\equiv& \frac{r_{\rm vir}(M)}{c_{\rm vir}} .
\end{eqnarray}
Finally the virial radius $r_{\rm vir} = r_{\rm vir}(M,z)$ and the
concentration parameter are given as
\begin{eqnarray}
\label{eq:def_rvir}
&& r_{\rm vir} \equiv \left(\frac{3M}
{4\pi \Delta_{\rm vir} \Omega_0 \rho_{\rm crit}}\right)^{1/3}
\approx \frac{1.69}{1+z} 
\left(\frac{\Delta_{\rm vir}\Omega_0}{18\pi^2}\right)^{-1/3}
\left(\frac{M}{10^{15}h^{-1}M_\odot}\right)^{1/3} , \\
&& \Delta_{\rm vir} \approx
\left\{
\begin{array}{ll}
18\pi^2 \Omega(z)^{-0.7} & \mbox{$(\lambda_0=0)$} \\
18\pi^2 \Omega(z)^{-0.6} & \mbox{$(\lambda_0=1-\Omega_0)$}
\end{array}
\right. , \\
&& c_{\rm vir}(M, z) = \frac{9}{1+z} \left(\frac{M} 
{1.5\times10^{13}h^{-1}M_{\odot}}\right)^{-0.13} .
\end{eqnarray}
The expression for $c_{\rm vir}$ is the fitting formula in LCDM (Lambda
Cold Dark Matter) by Bullock et al. (2001).

Then the corresponding mass profile and the circular velocity are
computed in a straightforward manner as
\begin{eqnarray}
\label{eq:mr_nfw}
M(r) = 4\pi \int_0^r \frac{\delta_c \rho_{\rm crit} r^2}
 {(r/r_{\rm s})(1+r/r_{\rm s})^2} dr
= 4\pi \delta_c \rho_{\rm crit} r_s^3 
\left[\ln\left(1+\frac{r}{r_s}\right) - \frac{r}{r+r_s}\right] ,
\end{eqnarray}
and
\begin{eqnarray}
V_c(r) = \sqrt{\frac{GM(r)}{r}}
= \sqrt{4\pi \delta_c \rho_{\rm crit} r_s^2 
\left[\frac{r_s}{r}\ln\left(1+\frac{r}{r_s}\right) 
- \frac{r_s}{r+r_s}\right]} .
\end{eqnarray}
The above results explicitly show that one can in principle compare the
universal density profile with the observational data without any
unknown free parameter. This is why the reported disagreement with the
mass profile reconstructed from weak/strong lensing (Tyson et al. 1998)
and the rotation curves of dwarf/low surface brightness galaxies (e.g.,
Moore et al. 1999; de Blok et al. 2001), if real, cannot be easily
accommodated in the context of the standard paradigm of cold dark
matter.

\subsection{X-ray gas density profile from the universal density profile
of dark halos:  origin of the isothermal $\beta$ model ?}

Another important application of the universal density profile of dark
halos may be found in X-ray emission from galaxy clusters.  A simple but
realistic approximation is the isothermal gas distribution in
hydrostatic equilibrium embedded in the NFW halo gravitational
potential. Actually the hydrostatic equilibrium equation:
\begin{eqnarray}
\frac{d p_{\rm gas}}{dr} = - \frac{GM(r)}{r^2}\rho_{\rm gas} ,
\end{eqnarray}
combined with the NFW profile and equation of state
\begin{eqnarray}
p_{\rm gas} = n_{\rm gas}kT_{\rm gas} 
= \frac{\rho_{\rm gas}}{\mu m_p}kT_{\rm gas} 
\end{eqnarray}
can be analytically integrated (Makino, Sasaki \& Suto 1998; Suto,
Sasaki \& Makino 1998) as
\begin{eqnarray}
\rho_{\rm gas}(r) = \rho_{\rm g0} ~ e^{-B} 
\left(1+\frac{r}{r_s}\right)^{Br_s/r} ,
\qquad B \equiv \frac{G\mu m_p}
{kT_{\rm gas}}4\pi \delta_c \rho_{\rm crit}r_s^2.
\end{eqnarray}
More surprising is the fact that the above model is very close to the
conventional $\beta$ model. Indeed we found the following fit as
illustrated in Figure 3:
\begin{eqnarray}
\rho_{\rm gas}(r) &\approx &
\frac{\rho_{g0}A}{[1+(r/r_{\rm c,eff})^2]^{3\beta_{\rm eff}/2}}, \\
A &\approx & - 0.178b + 0.982 = -0.013B + 0.982~(b \equiv 2B/27) , \cr
r_{\rm c,eff} &\approx& 0.22r_s , \qquad
\beta_{\rm eff} \approx  0.9b = 0.067B. \nonumber
\end{eqnarray}
\begin{figure}[h]
\begin{minipage}{5cm}
\leavevmode\epsfxsize=5cm \epsfbox{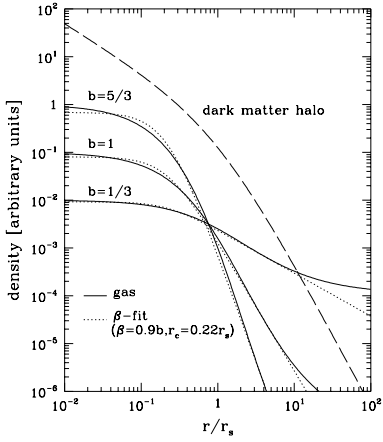}
\end{minipage}
\begin{minipage}{8cm}
\caption{Gas density profile (solid lines) 
expected from the universal density profile of dark matter halo
(dashed line) for $b=1/3$, $1$, and $5/3$. For comparison, the
best-fit $\beta$-models with $\beta=0.9b$ and $r_c=0.22r_s$ are
plotted in dotted lines (Makino, Sasaki, \& Suto 1998).
\label{fig:nfwbetamodel}}
\end{minipage}
\end{figure}

As far as I understand, this work is the first analytical model for the
cluster gas distribution obtained by consistently solving the Poisson
equation for the underlying halo mass distribution. The fact that the
resulting gas profile is quite similar to the isothermal $\beta$ model,
widely used as a good empirical model, is encouraging. On the other
hand, the expected core sizes of clusters seem to be significantly
smaller than the observational ones (Fig. 4).  In fact,
rotation curves of dwarf/LSB galaxies (Moore et al. 1999; de Blok et
al. 2001) and the mass profile reconstructed for the galaxy cluster
CL0024 (Tyson et al. 1998) also indicate the presence of the core,
rather than cusp, in the halo density profile. While this is still
controversial at this point, Spergel \& Steinhard (2000) among others
even proposed to abandon the collisionless nature of dark matter to
reconcile the reported discrepancy.

\begin{figure}[t]
\begin{minipage}{5cm}
\leavevmode\epsfxsize=5cm \epsfbox{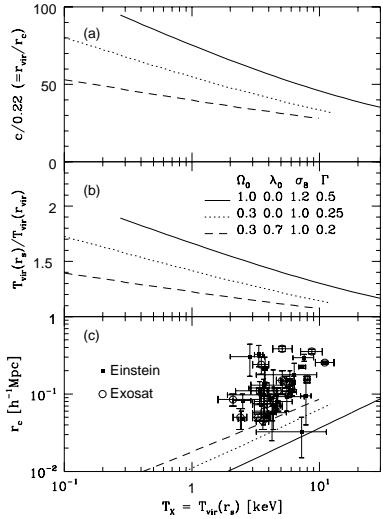}
\end{minipage}
\begin{minipage}{8cm}
\caption{Predicted properties of gas density distribution as functions 
of the X-ray cluster gas temperature $T_X=T_{\rm vir}(r_s)$.  (a)
Ratio of $r_{vir}$ and $r_c (\approx 0.22 r_s)$; (b) ratio of the
virial temperatures at $r_s$ and at $r_{\rm vir}$; (c) predicted sizes
of the effective core radius compared with the observed cluster data.
Solid, dotted and dashed lines indicate SCDM, OCDM, and LCDM models,
respectively.
\label{fig:rc}}
\end{minipage}
\end{figure}

\subsection{Collisional dark matter}

Since the cold dark matter model is known to successfully reproduce the
large-scale structure of the universe ($r \gg 1$Mpc), the possible
collision of dark matter should be effective only at scales below 1Mpc.
The required collision cross section $\sigma$ for a dark matter particle
of mass $m$ may be obtained by assuming
\begin{eqnarray}
(mn) \left(\frac{\sigma}{m}\right)l \approx 1
\quad {\rm for} 
~ mn \approx \rho_{\rm c, cluster} 
~ {\rm and} ~ l  \approx 1 {\rm Mpc}.
\end{eqnarray}
This may be rewritten as
\begin{eqnarray}
\left(\frac{\sigma}{m}\right) \approx 1.6 h^{-1} {\rm cm}^2/{\rm g}
\left(\frac{10^4 \rho_{\rm crit}}{\rho_{\rm c, cluster}}\right)
\left(\frac{1h^{-1}{\rm Mpc}}{l}\right)
\end{eqnarray}
or equivalently as
\begin{eqnarray}
\sigma \approx 2.5\times10^{-24} h^{-1} {\rm cm}^2
\left(\frac{m}{1{\rm GeV}}\right)
\left(\frac{10^4 \rho_{\rm crit}}{\rho_{\rm c, cluster}}\right)
\left(\frac{1h^{-1}{\rm Mpc}}{l}\right) .
\end{eqnarray}
The corresponding mean collision time-scale is
\begin{eqnarray}
\Delta t \approx \frac{l}{v}
\approx 10^9 
\left(\frac{1000{\rm km/s}}{v}\right)
\left(\frac{l}{1h^{-1}{\rm Mpc}}\right) h^{-1} {\rm year}.
\end{eqnarray}
The above rough estimate is encouraging because in the Hubble time
collisional dark matter with the cross section of an order $\sigma/m
\sim 1$cm$^2$/g will affect the central part of dark halos only while
keeping the larger-scale behavior intact.

More detailed study of the effect of collision should be made again with
numerical simulations. Yoshida et al. (1999) and Dav\'e et al. (2001)
conducted such simulations and found that dark matter models with
$\sigma/m \approx 1{\rm cm}^2/g$ lead to the formation of the softened
central core, rather than cusp, in cluster-sized halos.  The models,
however, simultaneously produce too spherical halos to be compatible
with the gravitational lensing observations (Miralda-Escude 2002).  In
addition, there are other indications to favor the presence of halos on
the basis of the gravitational lensing statistics (Molikawa \& Hattori
2001; Oguri et al. 2001; Takahashi \& Chiba 2001; Li \& Ostriker
2001). Furthermore, CL0024 seems to be a very complicated system and
thus the interpretation of its density profile is not straightforward
(Czoske et al. 2002).

\section{Beyond spherical description of dark halos}

Actually it is rather surprising that the fairly accurate scaling
relation applies after the spherical average despite the fact that the
departure from the spherical symmetry is quite visible in almost all
simulated halos (Fig. 1).  A more realistic modeling of
dark matter halos beyond the spherical approximation is important in
understanding various observed properties of galaxy clusters and
non-linear clustering (especially the high-order clustering statistics)
of dark matter in general.  In particular, the non-sphericity of dark
halos is supposed to play a central role in the X-ray morphologies of
clusters, in the cosmological parameter determination via the
Sunyaev-Zel'dovich effect and in the prediction of the cluster weak
lensing and the gravitational arc statistics (Bartelmann et al. 1998;
Meneghetti et al. 2000, 2001; Molikawa \& Hattori 2001, Oguri et al
2001).

Recently Jing \& Suto (2002) presented a detailed non-spherical modeling
of dark matter halos on the basis of a combined analysis of the
high-resolution halo simulations (12 halos with $N\sim 10^6$ particles
within their virial radius) and the large cosmological simulations (5
realizations with $N=512^3$ particles in a $100h^{-1}$Mpc boxsize).  The
density profiles of those simulated halos are well approximated by a
sequence of the concentric triaxial distribution with their axis
directions being fairly aligned.  They characterize the triaxial model
quantitatively by generalizing the universal density profile which has
previously been discussed only in the framework of the spherical model,
and obtain a series of practically useful fitting formulae in applying
the triaxial model; the mass and redshift dependence of the axis ratio,
the mean of the concentration parameter, and the probability
distribution functions of the the axis ratio and the concentration
parameter. Their triaxial description of the dark halos will be
particularly useful in predicting a variety of nonsphericity effects, to
a reasonably reliable degree, including the weak and strong lens
statistics, the orbital evolution of galactic satellites and triaxiality
of galactic halos, and the non-linear clustering of dark matter. Some
applications of the triaxial halo model is now in progress (Lee \& Suto
2002).
\begin{figure}[ht]
\vspace*{-0.5cm}
\begin{center}
 \leavevmode\epsfxsize=7.5cm \epsfbox{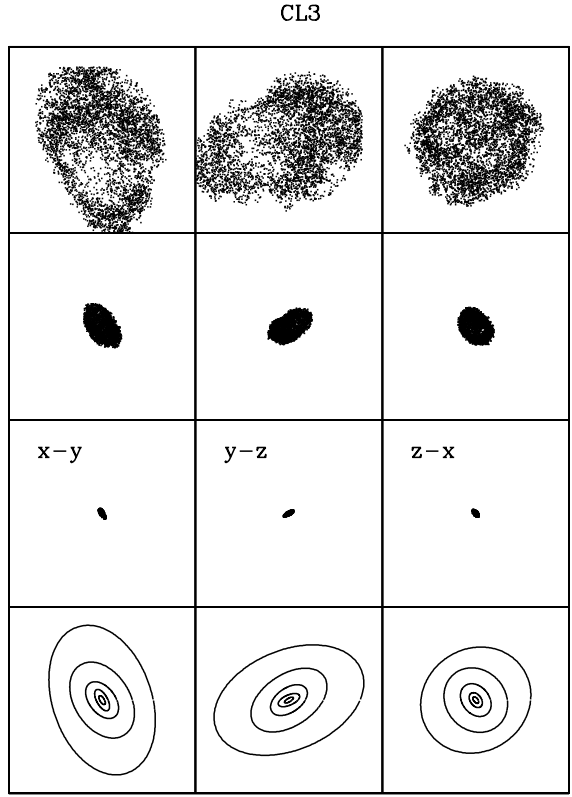} \caption{An
example of projected particle distribution for a cluster-size simulated
halo labelled as CL3.  The size of each box is $2r_{\rm vir}$ of the
halo, and particles in the isodensity shells with $\rho_s/\rho_{\rm
crit}=100$, $2500$, and $6.25\times 10^4$ are plotted on the $xy$, $yz$
and $zx$ planes (from left to right). The bottom panels show the
triaxial fits to five isodensity surfaces projected on those planes
(Jing \& Suto 2002).  \label{fig:project} }
\end{center} 
\end{figure}

\vspace*{-1cm}
\section{Clustering of dark halo on the light-cone}

All cosmological observations are carried out on a light-cone, the null
hypersurface of an observer at $z=0$, and not on any constant-time
hypersurface.  Thus clustering amplitude and shape of objects should
naturally evolve even {\it within} the survey volume of a given
observational catalogue. Unless restricting the objects at a narrow bin
of $z$ at the expense of the statistical significance, the proper
understanding of the data requires a theoretical model to take account
of the average over the light cone (Matsubara, Suto, \& Szapudi 1997;
Mataresse et al.  1997; Moscardini et al.  1998; Nakamura, Matsubara, \&
Suto 1998; Yamamoto \& Suto 1999; Suto et al. 1999).  We take account of
all relevant physical effects and construct an empirical model for the
two-point correlation functions of dark halos on the light-cone (Hamana,
Colombi, \& Suto 2001a; Hamana, Yoshida, Suto \& Evrard 2001b).

\begin{figure}[h]
  \begin{center}
 \leavevmode\epsfxsize=8.0cm \epsfbox{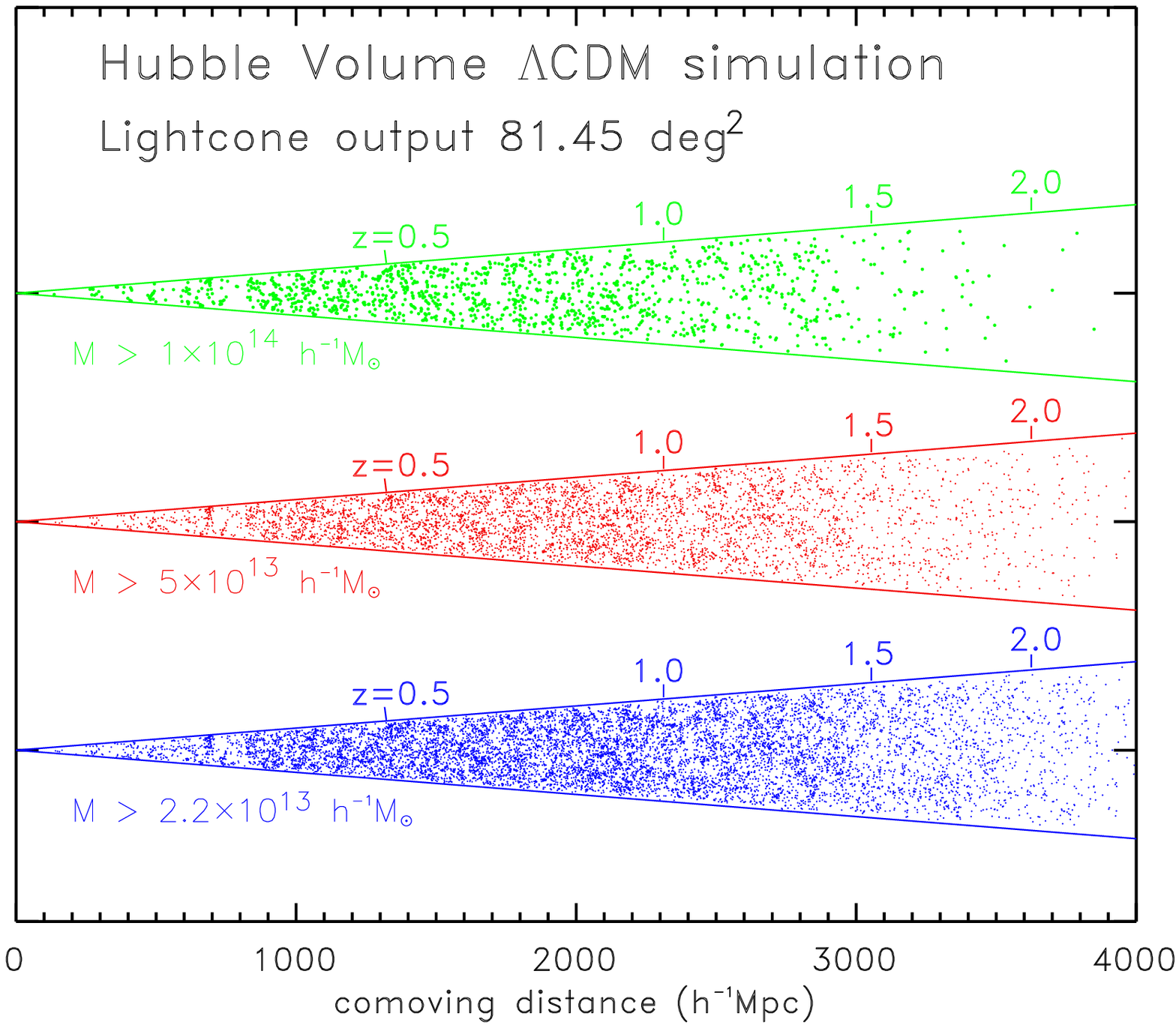}
\caption{Projected distribution of dark matter halos with $M_{\rm halo}
> M_{\rm min}$ in the light-cone output of the Hubble volume simulation;
$M_{\rm min}=1\times 10^{14}h^{-1}M_\odot$, $5\times
10^{13}h^{-1}M_\odot$ and $2.2\times 10^{13}h^{-1}M_\odot$ from top to
bottom.  } \label{fig:halos2d}
  \end{center}
\begin{center}
 \leavevmode\epsfxsize=7.5cm \epsfbox{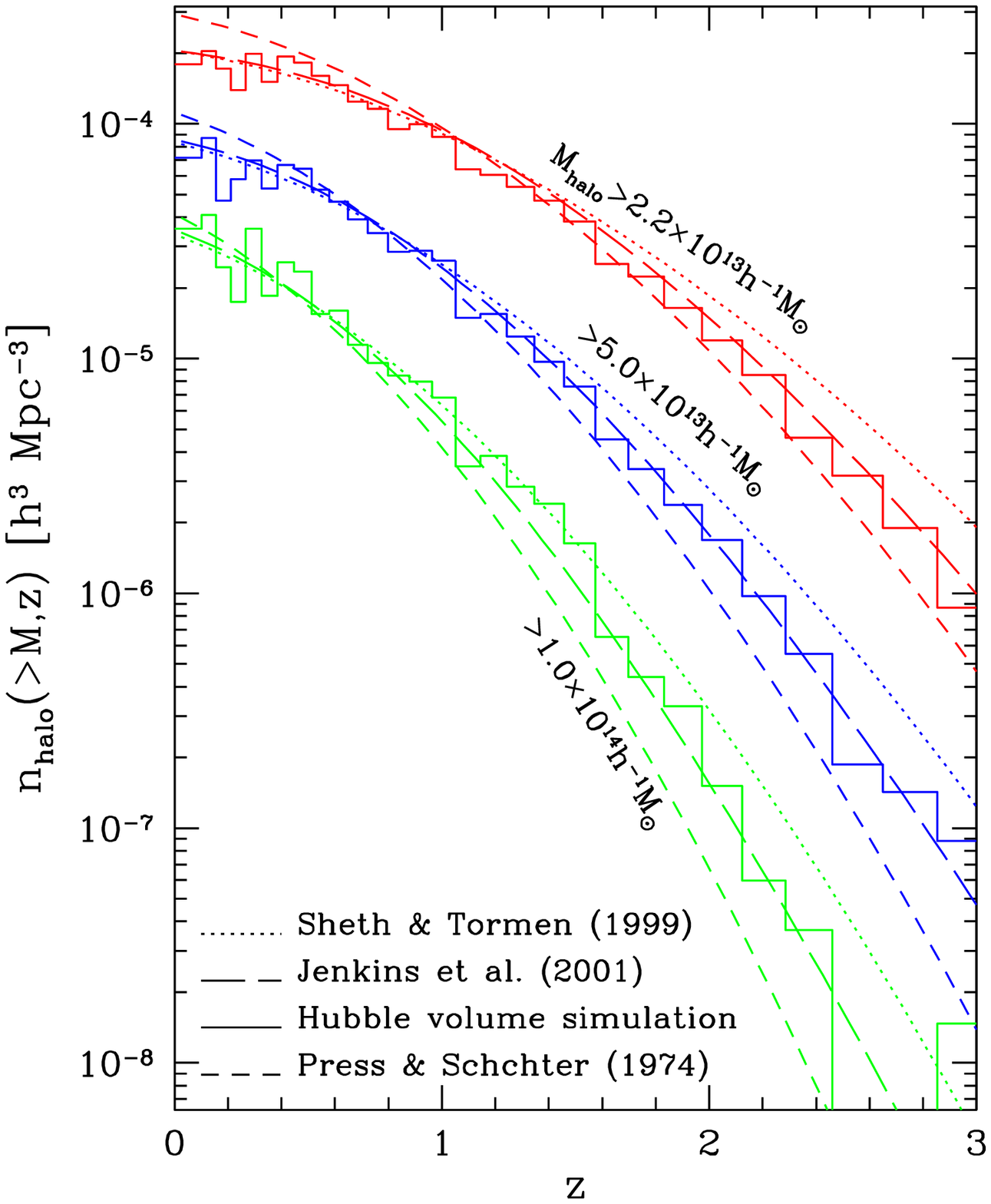}
\caption{Redshift distribution of the cumulative mass function of
 halos identified in the light-cone output of the Hubble volume
 simulation (histograms).  \label{fig:nz} }
\end{center} 
\end{figure}
\clearpage

Figure 6 plots the projected distribution of identified halos from
``light-cone output'' of the Hubble Volume $\Lambda$CDM simulation
(Evrard et al. 2002) with $\Omega_{\rm b}=0.04$, $\Omega_{\rm
CDM}=0.26$, $\sigma_{8}=0.9$, $\Omega_{\Lambda}=0.7$ and $h=0.7$.

 Figure 7 displays the redshift distribution of the cumulative mass
functions of the identified halos with $M>2.2\times 10^{13}
h^{-1}M_{\odot}$, $5.0\times 10^{13} h^{-1}M_{\odot}$, and $1.0\times
10^{14} h^{-1}M_{\odot}$. The total number of the corresponding halos at
$0<z<3$ is 21,090, 5,554 and 1,543, respectively.  For comparison, we
also plot the predictions on the basis of the Press-Schechter mass
function ({\it short-dashed lines}), the modified mass function by Sheth
\& Tormen (1999; {\it dotted lines}), and the fitting model by Jenkins
et al. (2001; {\it long-dashed lines}).  The Press-Schechter model
underpredicts the halo abundance at $z>1$, while the Sheth-Tormen model
overpredicts beyond $z\sim1.5$. This tendency is consistent with the
previous finding of Jenkins et al. (2001) that the Sheth-Tormen model
overestimates the number of halos when ln($\sigma^{-1}$) becomes large.

\begin{figure}[h]
\begin{center}
\leavevmode\epsfxsize=11cm \epsfbox{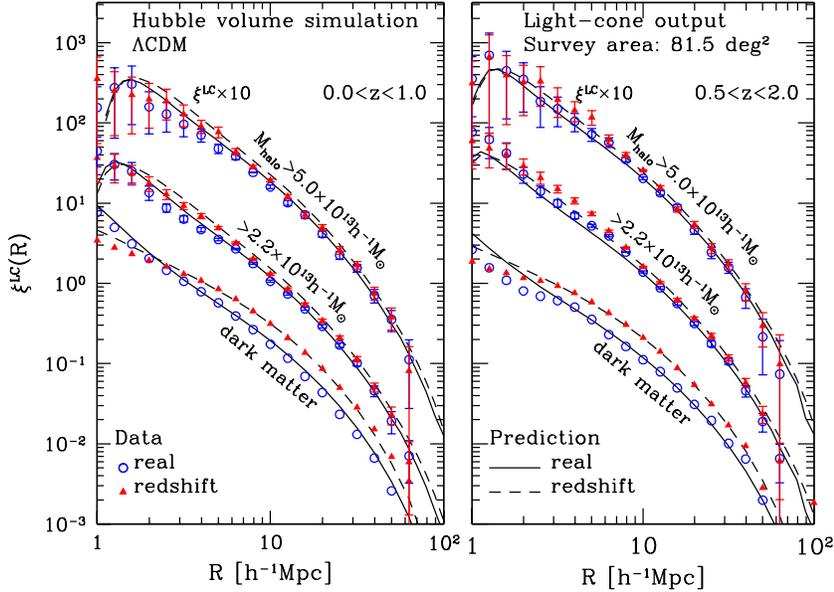}
\caption{Two-point correlation functions of halos on the light-cone;
simulation results (symbols; open circles and filled triangles for real
and redshift spaces, respectively) and our predictions (solid and dotted
lines for real and redshift spaces, respectively).  The error bars
denote the standard deviation computed from 200 random re-samplings of
the bootstrap method.  The amplitudes of $\xi^{LC}$ for $M_{\rm halo}\ge
5.0\times 10^{13}h^{-1} M_\odot$ are increased by an order of magnitude
for clarity.
\label{fig:haloxilc}}
 \end{center}
\end{figure}

Figure 8 compares our model predictions with the clustering of simulated
halos.  For the dark matter correlation functions, our model reproduces
the simulation data almost perfectly at $R>3h^{-1}$Mpc (see also Hamana
et al. 2001a). This scale corresponds to the mean particle separation of
this particular simulation, and thus the current simulation
systematically underestimates the real clustering below this scale
especially at $z>0.5$.  Our model and simulation data also show quite
good agreement for dark halos at scales larger than $5h^{-1}$Mpc. Below
that scale, they start to deviate slightly in a complicated fashion
depending on the mass of halo and the redshift range.  This discrepancy
may be ascribed to both the numerical limitations of the current
simulations and our rather simplified model for the halo biasing.
Nevertheless the clustering of {\it clusters} on scales below
$5h^{-1}$Mpc is difficult to determine observationally anyway, and our
model predictions differ from the simulation data only by $\sim 20$
percent at most. Therefore we conclude that in practice our empirical
model provides a successful description of halo clustering on the
light-cone.

\section{What's next ? : dark halos vs. galaxy clusters}

\begin{figure}[h]
\begin{center}
\leavevmode\epsfxsize=11cm \epsfbox{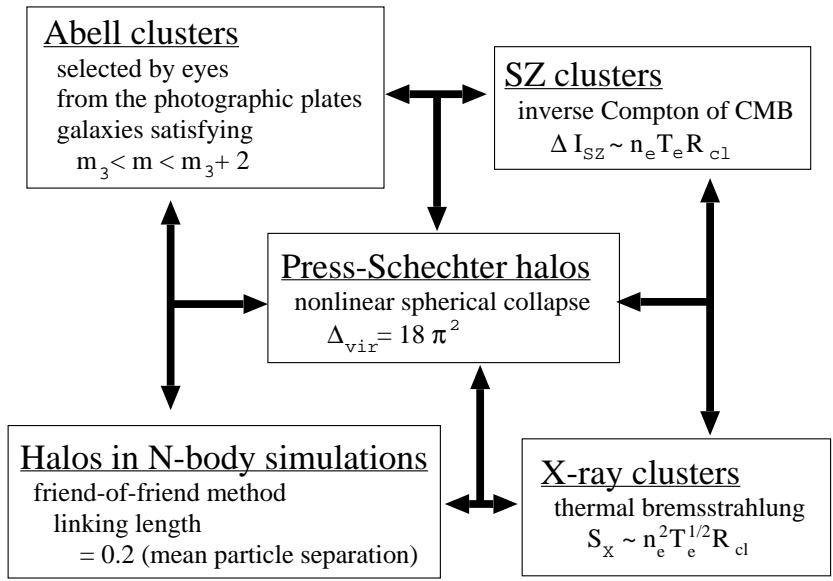}
\caption{Dark halos -- galaxy clusters connection. 
\label{fig:halocluster}}
\end{center}
\end{figure}
As I have briefly shown in the above sections, understanding of the
density profiles and clustering of dark halos has been significantly
advanced during last several years.  With those theoretical/empirical
successes in mind, the next natural question is how to apply them for
the description of {\it real} galaxy clusters. In principle this may be
fairly straightforward, but in reality the main obstacle is the lack of
the proper definition of clusters.  This point of view is clearly
illustrated in Figure 9 which summarizes the conventional definitions of
{\it clusters/halos} in various situations.  There are a wide range of
practical (and quite different!) definitions of dark halos and clusters
of galaxies. Of course they are closely related, but the one-to-one
correspondence is unlikely and nothing but a working hypothesis.  We we
need more quantitative justification or modification of the working
hypothesis in order to move on to {\it precision cosmology with
clusters}.
 
This problem has not been considered seriously probably because the
agreement between model predictions and available observations seems
already {\it satisfactory}. In fact, since current viable cosmological
models are specified by a set of many {\it adjustable} parameters, the
agreement does not necessarily justify the underlying assumption. Thus
it is dangerous to stop doubting the unjustified assumption because of
the (apparent) success.

\acknowledgements

I would like to thank S. Bowyer and C.-Y. Hwang for inviting me to this
well-organized and fruitful meeting.  The plots that I presented in
these proceedings are based on my collaborative work with A.E.Evrard,
T.Hamana, Y.P.Jing, N.Makino, S.Sasaki and N.Yoshida.  This research was
supported in part by the Grant-in-Aid for Scientific Research of JSPS
(12640231, 14102004).



\end{document}